\documentclass[conference]{IEEEtran}
\usepackage[top=0.7in, bottom=0.93in, left=0.625in, right=0.625in]{geometry} % ICC submission edas

\usepackage{cite}
\usepackage{amsmath,amssymb,amsfonts}
\usepackage{algorithmic}
\usepackage{graphicx}
\usepackage{textcomp}
\usepackage{xcolor}
\usepackage{array}
\usepackage{multirow}
\usepackage{enumitem}
\usepackage{makecell}
\usepackage{float}% better control of floating figures and tables
\usepackage{subfig}% for subfigures within figures
\usepackage{adjustbox}
\usepackage{booktabs}
\usepackage{longtable}
\usepackage{tabularx}
\usepackage{afterpage}
\usepackage{stfloats}

\usepackage{caption}
\newcolumntype{B}{>{\raggedright\arraybackslash}m{2 cm}}
\newcolumntype{D}{>{\centering\arraybackslash}m{3 cm}}
\newcolumntype{z}{>{\centering\arraybackslash}m{0.7 cm}}
\newcolumntype{L}{>{\centering\arraybackslash}m{1 cm}}

\newcolumntype{x}{>{\centering\arraybackslash}m{0.8 cm}}
\newcolumntype{e}{>{\centering\arraybackslash}m{1.0 cm}}
\newcolumntype{P}{>{\centering\arraybackslash}m{1.5 cm}}
\newcolumntype{?}{!{\vrule width 1.2pt}}

\def\BibTeX{{\rm B\kern-.05em{\sc i\kern-.025em b}\kern-.08em
		T\kern-.1667em\lower.7ex\hbox{E}\kern-.125emX}}

\usepackage{fancyhdr}
\fancypagestyle{firststyle}{\fancyhf{}

		\fancyhead[L]{ \small D. Shakya, M. Ying, T. S. Rappaport, P. Ma, I. Al-Wazani, Y. Wu, Y. Wang, D. Calin, H. Poddar, A. Bazzi, M. Chafii, Y. Xing, and A. Ghosh, ``Urban Outdoor Propagation Measurements and Channel Models at 6.75 GHz FR1(C) and 16.95 GHz FR3 Upper Mid-band Spectrum for 5G and 6G,"\textit{IEEE International Conference on Communications (ICC), }Montreal, Canada, Jun. 2025, pp. 1--6.}
	}
\usepackage{times}
\begin{document}
	\bstctlcite{IEEEtran:BSTcontrol}
	\title{Urban Outdoor Propagation Measurements and Channel Models at 6.75 GHz FR1(C) and 16.95 GHz FR3 Upper Mid-Band Spectrum for 5G and 6G}

	\author{
		\IEEEauthorblockN{Dipankar Shakya$^{\dagger1}$, Mingjun Ying$^{\dagger2}$, Theodore S. Rappaport$^{\dagger3}$, Peijie Ma$^\dagger$, Idris Al-Wazani$^\dagger$, Yanze Wu$^\dagger$,\\ Yanbo Wang$^\dagger$, Doru Calin$^*$, Hitesh Poddar$^\ddagger$, Ahmad Bazzi$^\S$$^\dagger$, Marwa Chafii$^\S$$^\dagger$, Yunchou Xing$^\P$, Amitava Ghosh$^\P$}
		
		\IEEEauthorblockA{
		$^\dagger$NYU WIRELESS, Tandon School of Engineering, New York University, USA\\
		$^*$MediaTek USA Inc., Warren, NJ, USA\\
		$^\ddagger$Sharp Laboratories of America (SLA), Vancouver, Washington, USA\\
		$^\S$Engineering Division, New York University Abu Dhabi, UAE\\
		$^\P$NOKIA Standards, Naperville, IL, USA\\
		\{$^1$dshakya, $^2$yingmingjun, $^3$tsr\}@nyu.edu
		}
		
		\thanks{This research is supported by the New York University (NYU) WIRELESS Industrial Affiliates Program.}
		}
	
	\maketitle
	
	\linespread{1.05}
	
	\thispagestyle{firststyle}
	
	\begin{abstract}
	Global allocations in the upper mid-band spectrum (4--24 GHz) necessitate a comprehensive exploration of the propagation behavior to meet the promise of coverage and capacity. This paper presents an extensive Urban Microcell (UMi) outdoor propagation measurement campaign at 6.75 GHz and 16.95 GHz conducted in Downtown Brooklyn, USA, using a 1 GHz bandwidth sliding correlation channel sounder over 40--880 m propagation distance, encompassing seven Line of Sight (LOS) and 13  Non-Line of Sight (NLOS) locations. Analysis of the path loss (PL) reveals lower directional and omnidirectional PL exponents compared to mmWave and sub-THz frequencies in the UMi environment, using the close-in (CI) free space PL (FSPL) model with a 1 m reference distance. Additionally, a decreasing trend in root mean square (RMS) delay spread (DS) and angular spread (AS) with increasing frequency was observed. The measured NLOS RMS DS and RMS AS mean values (as computed by 3GPP methods) are found to be consistently lower compared to 3GPP model predictions. Point-data tables with corresponding site-specific environmental information for all measured statistics at each TX-RX location are provided to support the models and results. The spatio-temporal statistics presented here offer valuable insights for the design of next-generation wireless systems and networks.
	\end{abstract}
	
	
	\section{Introduction}
	The upper mid-band spectrum is a major candidate for the deployment of cellular wireless services for 5G and 6G communications, with the FR1(C) (4--8 GHz) and  Frequency Range 3 (FR3, 7--24 GHz) bands showing promise for increased capacity and coverage \cite{Shakya2024ojcom,Kang2024OJCOM,Zhang2024CM,rappaport2012rws}. Recognizing the potential of the upper mid-band frequencies, telecommunications authorities across the world have made allocations in the upper mid-band spectrum for cellular services \cite{Shakya2024ojcom}. Particularly for outdoor urban environments, understanding the physical channel is crucial to drive infrastructure deployments and facilitate effective air-interface design \cite{rappaport2017investigation,sun2014icc}. The evaluation of path loss, multipath delay spread and angular spreads are critical for parametrizing the wireless channel models and the numerology for cellular air-interface design, such as sub-carrier spacing, cyclic prefix length for waveform design, or beamforming and beam management\cite{Zaidi2016ICM,Zhang2024CM}. However, there are limited studies analyzing the radio propagation characteristics in the upper mid-band spectrum in the urban microcell (UMi) environment.
	
	Zhang \textit{et al.} in \cite{Zhang2024CM} presented observations for coverage and network design, and showed lower measured RMS delay spread (DS) than predicted by 3GPP models at 6.5 GHz and 15 GHz in an outdoor campus. The observations were based on measurements conducted in \cite{Miao2023JSAC} using PN sequence transmitted between a signal generator and spectrum analyzer pair with a rooftop TX at 12.5 m height and human height RX. 200 MHz BW was used at 6.5 GHz and 1 GHz BW at 15 GHz. Using a close-in (CI) FSPL model with a 1 m reference distance, the path loss exponent (PLE) in LOS was obtained as 2.07 and 2.01 at 6.5 GHz and 15 GHz, respectively, while NLOS PLEs of 2.38 and 2.59 were observed at 6.5 GHz and 15 GHz, respectively \cite{Zhang2024CM, Miao2023JSAC}. 
	Investigation of the channel characteristics at 10 GHz with 250 MHz BW in a street-canyon in Berlin with a 3.5 m high TX antenna and 1.5 m RX antenna height revealed PLE of 1.6 in LOS and 2.1 in NLOS using the CI FSPL model with 1 m reference distance \cite{peter2016sensors}. Additionally, the mean RMS DS was reported as 21.9 ns in LOS and 37.7 ns in NLOS \cite{peter2016sensors}. 
	Wideband urban channel measurements were carried out around the University of Southern California stepping the carrier frequency from 3 to 18 GHz in 500 MHz steps using a 1 GHz BW frequency-hopped multi-band channel sounder \cite{Kristem2018twc}. Using the non-physically based floating intercept (FI) \cite{sun2016propagation} PL model, the LOS $\alpha$ exponent was reported to vary between 1.5 to 2 (with a non-physical offset $\beta$ between 48.2 and 55.15 dB) and in NLOS, the $\alpha$ was observed between 2.2 and 4.3 (with $\beta$ having an enormous four orders of magnitude variation between 22 and 63 dB). The LOS RMS DS was observed between 25-50 ns and 30-140 ns in NLOS for the UMi measurements applying a threshold of 6 dB above noise floor. In LOS, the mean RMS DS was observed to decrease from 46 ns to 29 ns from 3 to 18 GHz \cite{Kristem2018twc}. 

	Channel characterization at 10.1 GHz with a vector network analyzer spanning 500 MHz BW in outdoor campus of the University of Oulu, resulted in PLEs of 2.0 in LOS and 3.0 in NLOS using the CI FSPL model and a 1 m reference distance \cite{Roivainen2017tap}. Mean RMS DS was observed to be 21.45 ns in LOS and 41.29 in NLOS (using 20 dB dynamic range). Additionally, mean angular spread of arrival (ASA) was reported to be 32.6$^\circ$ in LOS and 59.7$^\circ$ NLOS, whereas mean angular spread of departure (ASD) was reported to be 14.9$^\circ$ in LOS and 21.1$^\circ$ in NLOS. Elevation spreads for arrival and departure were reported as 7$^\circ$ and 6.8$^\circ$ in LOS and 13$^\circ$ and 8$^\circ$ in NOS, respectively \cite{Roivainen2017tap}. 
	
	This paper presents comprehensive models for path loss, RMS DS, and AS, and compares these statistics with existing 3GPP models and existing results at higher mmWave frequencies. The statistics are generated from an intense measurement campaign conducted around the NYU MetroTech Commons in Brooklyn, New York, during Spring 2024, using a 1 GHz bandwidth sliding correlation channel sounder encompassing seven LOS and 13 NLOS TX-RX location pairs \cite{Shakya2024ojcom}. The channel measurements reported in this paper corroborate the findings in \cite{Zaidi2016ICM} suggesting 3GPP overestimates the DS in NLOS UMi environments at upper mid-band frequencies and that minor modifications may be needed in terms of numerology compared to 5G NR. Moreover, the lower measured RMS ASA may suggest modifications for beam management parameters related to beam alignment accuracy, optimal beamwidth selection, and dynamic adaptation of the beam management process for a higher number of antenna elements at both gNB and UE \cite{Zhang2024CM,Roivainen2017tap}. 
	The remainder of the paper is organized starting with the channel sounder system details in Section II. A description of the measurement environment and procedures for propagation measurements at each location are noted in Section III. Section IV presents the large-scale path loss behavior. RMS DS and AS statistics are given in Section V, before concluding.        

    \renewcommand{\arraystretch}{1}
	\begin{table}[htbp]
		\centering
		\caption{Upper Mid-Band Channel Sounder System\cite{Shakya2024ojcom}}
		%\vspace{-1 em}
		\begin{tabular}{|p{3.2 cm}|>{\centering\arraybackslash}p{2 cm}|>{\centering\arraybackslash}p{2 cm}|}
			\hline
			\raggedleft{\textbf{Carrier Frequency }} & \textbf{6.75 GHz} & \textbf{16.95 GHz} \\
			\hline
			\raggedleft\textbf{Free Space PL at 1m reference distance} & \vfil 49 dB & \vfil 57 dB \\
			\hline
			\raggedleft \textbf{Baseband signal} & \multicolumn{2}{c|}{11th order PN sequence (2047 chips)} \\
			\hline
			\raggedleft \textbf{TX PN Code Chip Rate} & \multicolumn{2}{c|}{500 Mcps} \\
			\hline
			%\raggedleft \textbf{TX PN Code Chip Width} & \multicolumn{2}{c|}{2.0 ns} \\
			%\hline
			\raggedleft \textbf{RX PN Code Chip Rate} & \multicolumn{2}{c|}{499.9375 Mcps} \\
			\hline
			\raggedleft \textbf{Slide factor} & \multicolumn{2}{c|}{8000} \\
			\hline
			\raggedleft \textbf{RF BW (Null-to-null)} & \multicolumn{2}{c|}{1 GHz} \\
			\hline
			\raggedleft \textbf{TX/RX Antenna Gain} & 15 dBi & 20 dBi \\
			\hline
			\raggedleft \textbf{TX/RX Ant. HPBW (Az/El)} & \vfil 30$^{\circ}$ / 30$^{\circ}$ & \vfil 15$^{\circ}$ / 15$^{\circ}$ \\
			%investigate HPBW for 73 GHz
			\hline
			\raggedleft \textbf{TX/RX Antenna Height} & \multicolumn{2}{c|}{4 m / 1.5 m} \\
			\hline
			%\raggedleft\textbf{XPD} & 35 dB & 38 dB \\
			%\hline
			%\raggedleft\textbf{Max EIRP} & 44 dBm & 46.5 dBm \\
			%\hline
			\raggedleft\textbf{Max EIRP used} & \multicolumn{2}{c|}{31 dBm} \\
			\hline
			\raggedleft \textbf{Max Measurable Path Loss (at 5 dB SNR)} & \vfil 142.7 dB & \vfil 143.8 dB \\
			\hline
			\raggedleft \textbf{TX Polarization} & \multicolumn{2}{c|}{Vertical (V)\,/\,Horizontal (H)} \\
			\hline
			\raggedleft \textbf{RX Polarization} & \multicolumn{2}{c|}{Vertical/Horizontal} \\
			\hline
		\end{tabular}%
		\label{tab:SysParams}%
		\vspace*{-1\baselineskip}
	\end{table}
	\renewcommand{\arraystretch}{1}

	\section{Wideband FR1C/FR3 Sliding Correlation Channel Sounder}
	A sliding correlation based channel sounder was employed for capturing the time-domain channel impulse response. The sliding correlation operation uses pseudo-random noise (PN) sequences transmitted between the channel sounder TX and RX, as described in \cite{Shakya2021tcas,Shakya2024ojcom, Ju2019icc}. The main system features are summarized in Table \ref{tab:SysParams}.

    The Mini-Circuits dual-band RF unit is mounted on a mechanically rotatable gimbal with a one-degree angular resolution. The measurement system has a link margin of 155.6 dB at 6.75 GHz and 159.2 dB at 16.95 GHz\cite{Shakya2024ojcom,Ted2025icc,Shakya2025wcnc}. A maximum EIRP of 31 dBm is used during field measurements, offering a practical link margin of 142.7 dB at 6.75 GHz and 143.8 dB at 16.95 GHz.
	
	\subsection{Calibration}
	Channel sounder calibration is accomplished in three steps: (a) system linearity and power calibration, (b) time calibration, and (c) spatial calibration to achieve accurate capture of multipath component (MPC) power, delay, and directions, as detailed in \cite{Shakya2024ojcom, Shakya2024gc}.

	Daily startup calibration and recalibration at the end of each day verified system linearity and power variations typically within 0.2 dB \cite{Shakya2024ojcom}. For time calibration, a patent-pending synchronization method \cite{Shakya2023gc} is implemented using precision time protocol (PTP) implemented over a Wi-Fi side link to relay a synchronization pulse between Rubidium clocks at the TX and RX throughout the day. A successive drift correction eliminated any remnant drift through repeated observation of a reference MPC delay \cite{Shakya2024ojcom,Shakya2023gc}.
	
	For spatial calibration, with the TX and RX at their specified locations, the TX and RX horns were horizontally leveled with a leveling tool and 0$^\circ$ spatial reference for both systems was set to the geographical North.

	\section{Measurement Environment and Procedure}
\begin{figure}[!t]
		\centering%
		\subfloat[]{%
			\centering
			\includegraphics[width=87mm]{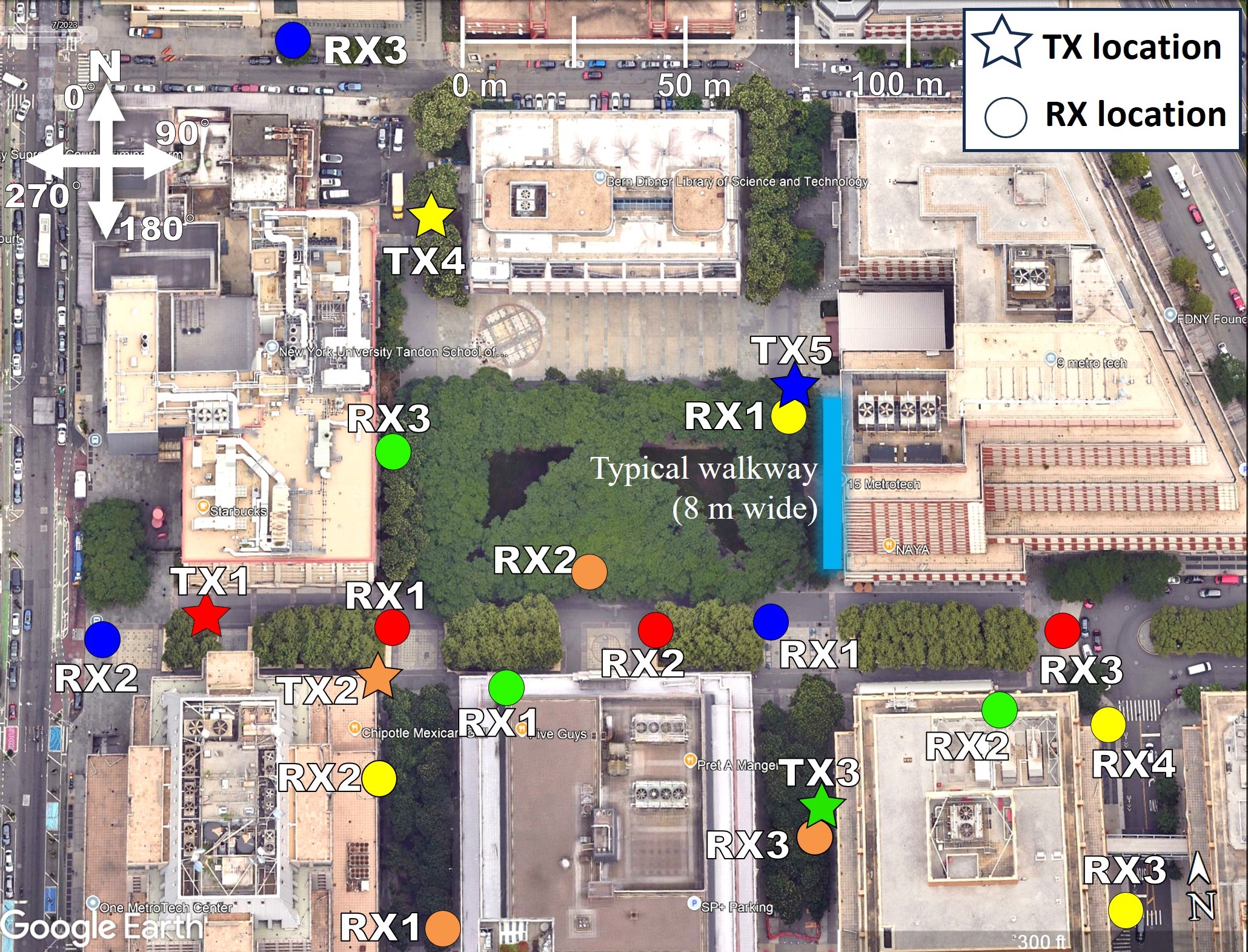}
		}\\
		\vspace{-5 pt}
		\subfloat[]{%
			\centering
			\includegraphics[width=87mm]{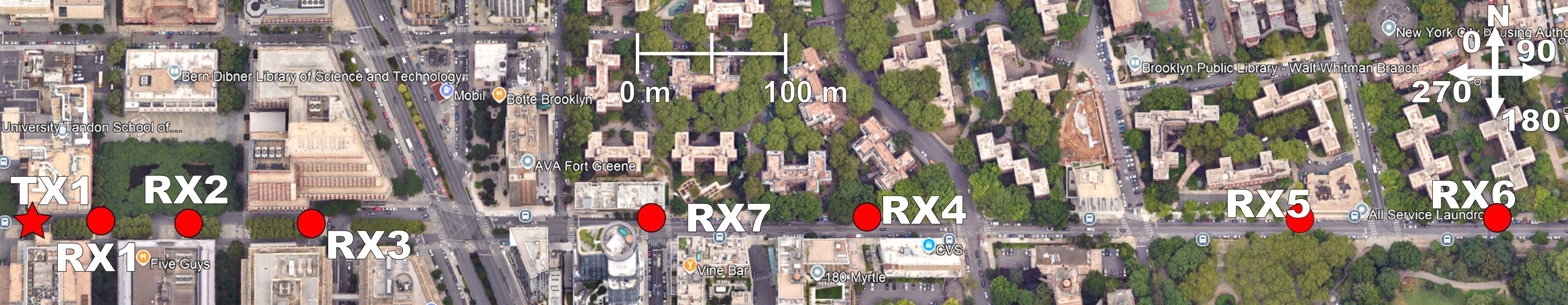}
		}%
		\\%[2.6mm]
		\caption{6.75 GHz FR1(C) \& 16.95 GHz FR3 campaign at the MetoTech Commons in Brooklyn, NY: (a) Open square of around 200 m. Typical walkways are 8 m wide (one walkway highlighted in sky-blue). (b) Long range measurements for TX1 up to 1 km.}
		\label{fig:ODMap}
		\vspace{-5 pt}
	\end{figure}

	The intense measurement campaign was conducted around the NYU Tandon School of Engineering campus in MetroTech, Brooklyn, NY and along a 1 km stretch on Myrtle Avenue. TX and RX locations are indicated in Fig. \ref{fig:ODMap} (a) and (b). TX-RX location pairs measured within the $\sim$180 m open-square configuration of the MetroTech Commons are shown in Fig. \ref{fig:ODMap} (a). 
	Additionally, Fig. \ref{fig:ODMap} (b) shows TX-RX location pairs along Myrtle Avenue, beyond MetroTech Commons, with separation distances ranging from 424 to 880 meters, representing a typical street path for a moving vehicle.
    The outdoor UMi multipath statistics were derived by analyzing propagation delay, power, and directions for all MPCs at each TX-RX location, following processes outlined in Table \ref{tab:sweeps}\cite{Shakya2024ojcom}.	
	
	\renewcommand{\arraystretch}{1.05}
	\begin{table}[!t]
		\centering
		\caption{TX/RX Elevation angles for different RX azimuthal sweeps for a fixed TX azimuth angle at each TX-RX location \cite{Shakya2024ojcom}}
		% \vspace{-5 pt}
		\begin{tabular}{|p{0.75 cm}|p{2.3 cm}|p{4.3 cm}|}
			\hline
			\textbf{Sweep\#} & \textbf{TX elevation} &\textbf{RX elevation} \\
			\hline
			\textbf{1}     & TX is kept at \textit{boresight}$^{\mathsection}$ elevation& RX is kept at \textit{boresight} elevation. RX is then swept 360$^{\circ}$ in the azimuth plane in HPBW steps. \\
			\hline
			\textbf{2}     & TX is kept at \textit{boresight} elevation& RX is \textit{tilted down by one HPBW} (30$^{\circ}$ at 6.75 GHz/15$^{\circ}$ at 16.95 GHz). RX is then swept 360$^{\circ}$ in the azimuth plane in HPBW steps. \\
			\hline
			\textbf{3}     &TX is kept at \textit{boresight} elevation& RX is \textit{tilted up by one HPBW} (30$^{\circ}$ at 6.75 GHz/15$^{\circ}$ at 16.95 GHz). RX is then swept 360$^{\circ}$ in the azimuth plane in HPBW steps. \\
			\hline
			\textbf{4}     & TX is \textit{tilted down by one HPBW} (30$^{\circ}$ at 6.75 GHz/15$^{\circ}$ at 16.95 GHz).& RX is kept at boresight elevation. RX is then swept 360$^{\circ}$ in the azimuth plane in HPBW steps. \\
			\hline
			\textbf{5}     & TX is \textit{tilted down by one HPBW} (30$^{\circ}$ at 6.75 GHz/15$^{\circ}$ at 16.95 GHz).& RX is \textit{tilted down by HPBW} (30$^{\circ}$ at 6.75 GHz/15$^{\circ}$ at 16.95 GHz). RX is then swept 360$^{\circ}$ in the azimuth plane in HPBW steps. \\
			\hline
            \multicolumn{3}{l}{\footnotesize $\mathsection$ \textit{As taught in \cite{Rappaport2015tc}, boresight elevation means small downtilt/uptilt.}}\\
            \multicolumn{3}{l}{\footnotesize \textit{for each TX/RX location.}}
		\end{tabular}%
		\label{tab:sweeps}%
		\vspace*{-1\baselineskip}
	\end{table}%
	\renewcommand{\arraystretch}{1.0}
    \vspace{-10 pt}
	
	\section{Large Scale Path Loss Models}%\vspace{-10 pt}
	The PL in directional and omnidirectional contexts are evaluated using a close-in (CI) FSPL model with a 1 m reference distance, \eqref{eq:CI}. The CI model describes PL with a single parameter, the PLE (n), referring to a free-space reference distance, $d_0$, as demonstrated in Appendix A of \cite{maccartney2015ia}.
    \vspace{-5 pt}
	{\small
	\begin{equation}
		\label{eq:CI}
		\begin{split}
			PL^{CI}(f_c,d_{\text{3D}})\;\text{[dB]} &= \text{FSPL}(f_c, 1 m)+ 10n\log_{10}\left( \dfrac{d_{3D}}{d_{0}} \right)+\chi_{\sigma},\\
			\text{FSPL}(f_c,1 m) &= 32.4 + 	20\log_{10}\left(\dfrac{f_c}{1\;\text{GHz}}\right),
		\end{split}
	\end{equation}}
	\noindent where FSPL$(f_c, 1 \;\text{m})$ is obtained for carrier frequency $f_c$ GHz at 1 m, $n$ is the PLE, and $\chi_{\sigma}$ is large-scale shadow fading (zero-mean Gaussian r.v. with s.d. $\sigma^{CI}$ in dB) \cite{Rappaport2015tc}.
	
		\renewcommand{\arraystretch}{1.2}
	\begin{table*}[!t]
		\centering
		%\color{blue}
		\caption{Directional and Omnidirectional CI FSPL 1 m reference distance model statistics for 6.75 and 16.95 GHz FR1(C) and FR3 UMi campaigns, as compared to 28, 73 and 142 GHz campaigns\cite{Shakya2024ojcom,Xing2021icl}}
		% \vspace{-1 em}
        \resizebox{0.9\textwidth}{!}{
		\begin{tabular}{|@{}D@{}|L|L|z|z|z|z|z|z|z|z|z|z|}
			\cline{1-13}
			\multirow{3}{2.1 cm}{\vfil \centering \textbf{Campaign}} & \multirow{3}{1 cm}{\vfil \centering \textbf{Distance (m)}} & \multirow{3}{1 cm}{\centering \textbf{Antenna HPBW (TX/RX)}} & \multicolumn{6}{c|}{\textbf{Directional path loss}} & \multicolumn{4}{c|}{\textbf{Omni path loss}} \\
			\cline{4-13}    \multicolumn{1}{|c|}{} & \multicolumn{1}{c|}{ } & \multicolumn{1}{c|}{ } & \multicolumn{2}{c|}{LOS} & \multicolumn{2}{c|}{NLOS Best} & \multicolumn{2}{c|}{NLOS} & \multicolumn{2}{c|}{LOS} & \multicolumn{2}{c|}{NLOS}  \\
			\cline{4-13}    \multicolumn{1}{|c|}{} & \multicolumn{1}{c|}{} & \multicolumn{1}{c|}{} & \multicolumn{1}{z|}{n} & \multicolumn{1}{z|}{$\sigma$ (dB)} & \multicolumn{1}{z|}{n} & \multicolumn{1}{z|}{$\sigma$ (dB)} & \multicolumn{1}{z|}{n} & \multicolumn{1}{z|}{$\sigma$ (dB)} & \multicolumn{1}{z|}{n} & \multicolumn{1}{z|}{$\sigma$ (dB)} & \multicolumn{1}{z|}{n} & \multicolumn{1}{z|}{$\sigma$ (dB)}\\
			\cline{1-13}
			\centering \textbf{6.75 GHz (This work)} & 40-1000 & (30$^{\circ}$/30$^{\circ}$) & 1.89  & 2.05  & 2.68  & 6.5  & 3.25  & 12.25  & 1.79  & 2.57  & 2.56  & 6.53 \\
			\cline{1-13}
			\centering \textbf{6.75 GHz (3GPP)} & 10-600 & -- & --  & --  & --  & --  & --  & --  & 2.19  & 4  & 3.19 & 8.2 \\
			\cline{1-13}
			\centering \textbf{16.95 GHz (This work)} & 40-1000  & (15$^{\circ}$/15$^{\circ}$) & 1.97  & 3.41  & 2.74  & 10.29  & 3.51  & 14.02 & 1.85  & 4.05  & 2.59  & 8.78 \\
			\cline{1-13}
			\centering \textbf{16.95 GHz (3GPP)} & 10-339 & -- & --  & --  & --  & --  & --  & -- & 2.1  & 4  & 3.19  & 8.2 \\
			\cline{1-13}
			\centering \textbf{28 GHz \cite{Shakya2024TAP}} & 31-187  & (11$^{\circ}$/11$^{\circ}$) & 2.27  & 8.15  & 3.79  & 7.97  & 4.13  & 10.11 & 2.02  & 8.98  & 3.56  & 8.91 \\
			\cline{1-13}
			\centering \textbf{73 GHz \cite{Shakya2024TAP}} & 21-170  & (7$^{\circ}$/15$^{\circ}$) & 1.92  & 1.75  & 2.98  & 9.36  & 3.73  & 11.47 & 1.87  & 1.64  & 2.79  & 8.56 \\
			\cline{1-13}
			\centering \textbf{142 GHz \cite{Xing2021icl}} & 21-117  & (8$^{\circ}$/8$^{\circ}$) & 2.02  & 2.71  & 3.07  & 8.71  & 3.48  & 9.53 & 1.96  & 2.63  & 2.92  & 8.28 \\
			\cline{1-13}
    		\end{tabular}}
		\label{tab:PLEs}%
		%\vspace*{-1\baselineskip}
		\vspace{-10 pt}
	\end{table*}%

	\renewcommand{\arraystretch}{1}
	
	\begin{figure}[!t]
		\centering%
		\subfloat[]{%
			\centering
			\includegraphics[width=90mm]{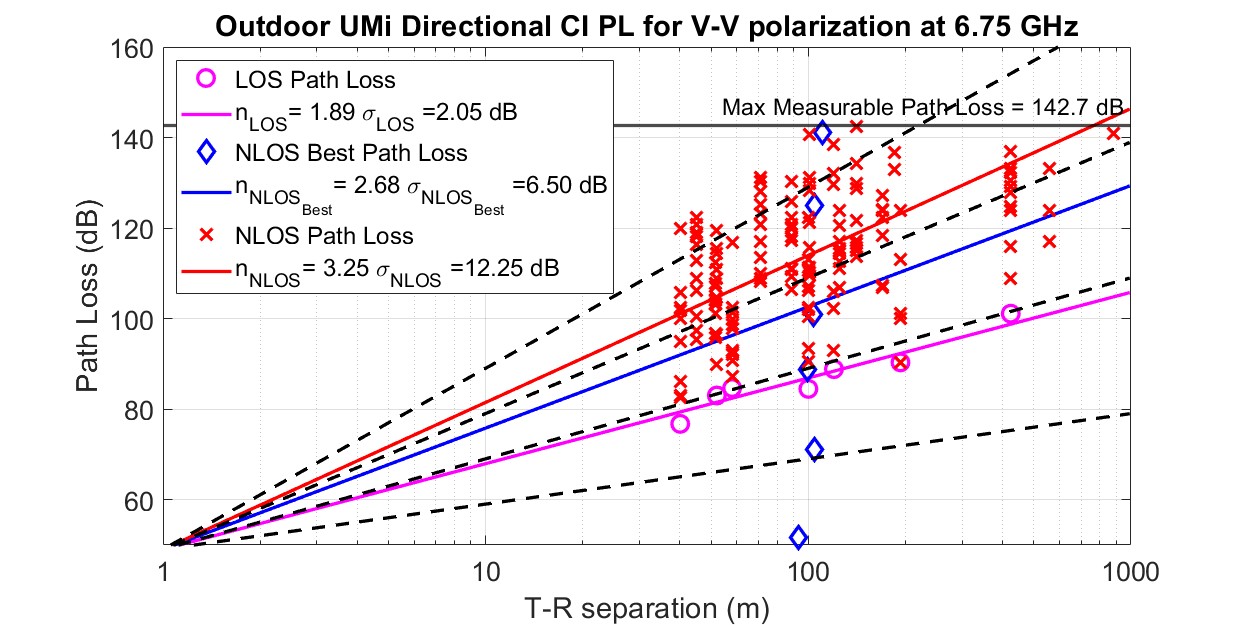}
		}\\
		\vspace{-5 pt}
		\subfloat[]{%
			\centering
			\includegraphics[width=90mm]{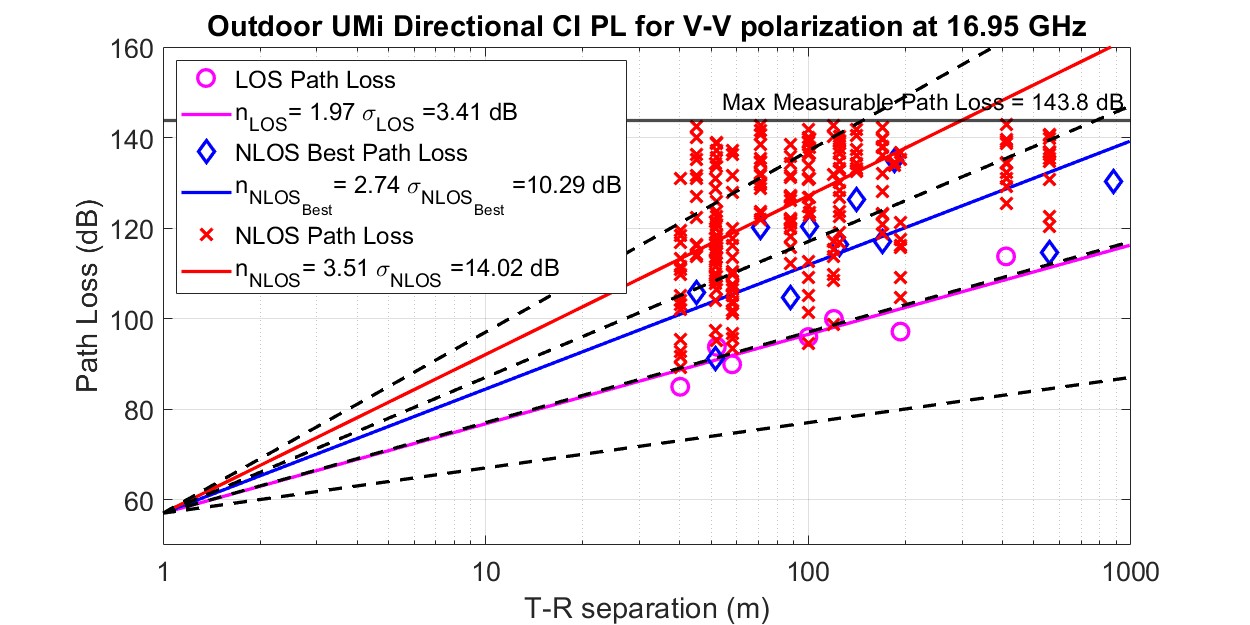}
		}%
		\\%[2.6mm]
		\caption{UMi directional CI FSPL models with a 1 m reference distance and scatter plot for V-V polarization at: (a) 6.75 GHz FR1(C); (b) 16.95 GHz FR3. [T-R separation: 40--880 m]}
		\label{fig:DirPL}
		\vspace{-15 pt}
	\end{figure}
	
\begin{figure}
		\centering%
		\subfloat[]{%
			\centering
			\includegraphics[width=85mm, height=40mm]{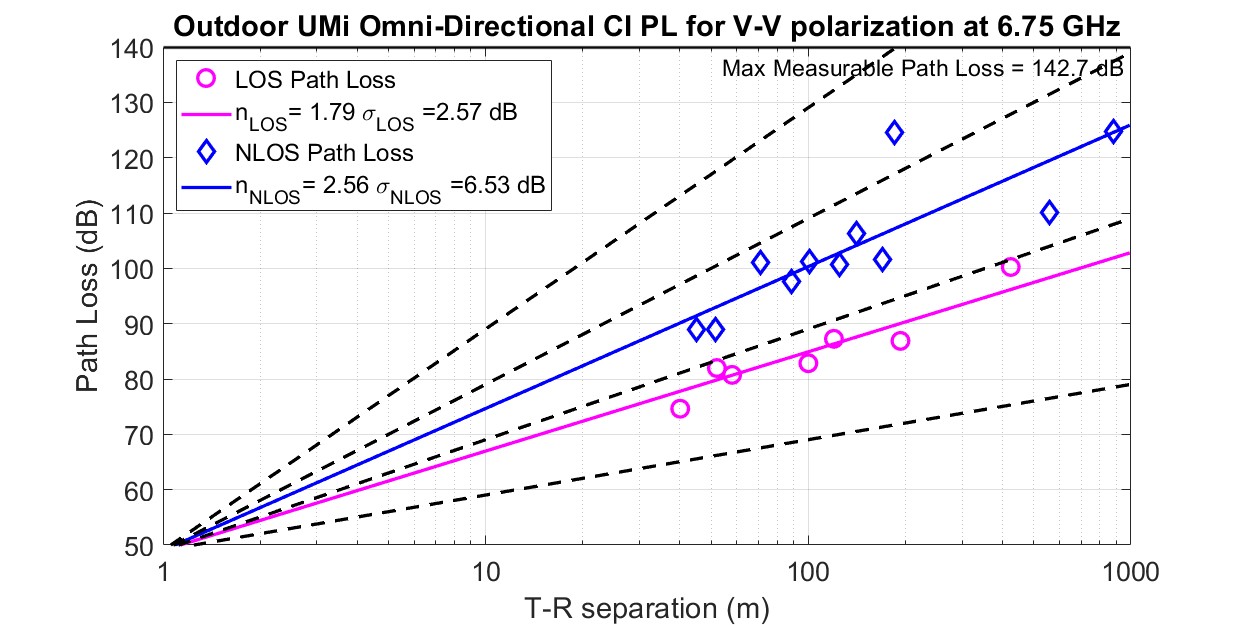}
		}\\
		\vspace{-10 pt}
		\subfloat[]{%
			\centering
			\includegraphics[width=85mm, height=40mm]{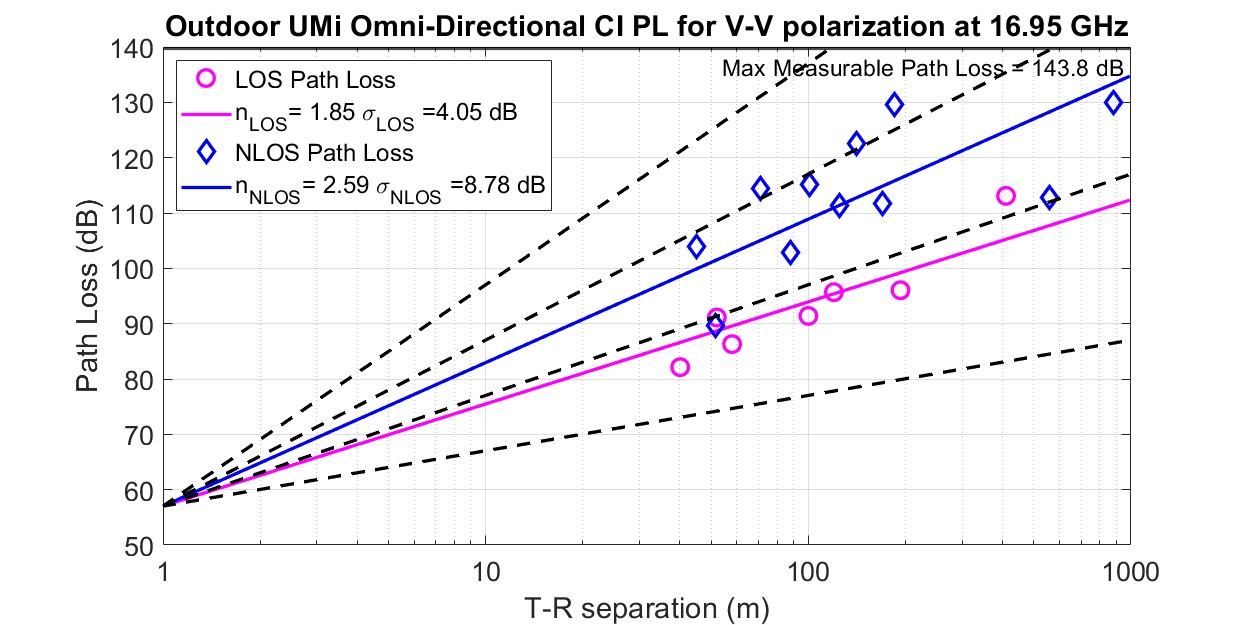}
		}%
		\\%[2.6mm]
		\caption{UMi omnidirectional CI FSPL models with a 1 m reference distance and scatter plot for V-V polarization at: (a) 6.75 GHz FR1(C); (b) 16.95 GHz FR3. [T-R separation: 40--880 m]}
		\label{fig:OmniPL}
		\vspace{-15 pt}
	\end{figure}

	Fig. \ref{fig:DirPL} presents the directional CI FSPL models with a 1 m reference distance and scatter plots for V-V polarization in the UMi environment at 6.75 GHz and 16.95 GHz. The measurements span T-R separations from 40 to 1000 m. Three scenarios are depicted: LOS, NLOS$_{\text{Best}}$, and NLOS. The NLOS$_{\text{Best}}$ PL is calculated for the RX pointing in the direction with the highest received power. At 6.75 GHz, the PLEs for LOS, NLOS$_{\text{Best}}$, and NLOS are 1.89, 2.68, and 3.25, respectively. %, with corresponding shadow fading standard deviations of 2.05 dB, 6.5 dB, and 12.25 dB. 
	At 16.95 GHz, the PLEs are higher at 1.97, 2.74, and 3.51, respectively. Fig. \ref{fig:OmniPL} illustrates the omni CI FSPL models with a 1 m reference distance and scatter plots in the same UMi environment and frequency bands. For the 6.75 GHz band, the LOS PLE is 1.79, while the NLOS PLE is 2.56. At 16.95 GHz, the LOS PLE slightly increases to 1.85, and the NLOS PLE is 2.59.
	
	As shown in Table \ref{tab:PLEs}, the measured PL results are compared with 3GPP models for mid-band, as well as past measurements at higher frequencies (28, 73, and 142 GHz). The measured PLEs at 6.75 and 16.95 GHz are generally lower than at higher frequencies, particularly for NLOS, indicating less severe path loss in the FR3 bands than at higher frequencies. 

	\section{Spatio-temporal Statistics}
	
	\subsection{Delay Spread Measurement Results}\vspace{-4 pt}
	The RMS DS quantifies the temporal dispersion of multipath components in a wideband channel. The RMS DS was evaluated using a power threshold of 25 dB below the PDP maxima or 5 dB above the noise floor \cite{Ying2025tcom}.
	Table \ref{tab:RMS_DS} presents the RMS DS characteristics from 6.75 GHz to 142 GHz for both LOS and NLOS conditions, comparing results with existing 3GPP models and higher frequency results.
	For directional measurements at 6.75 GHz, the mean RMS DS is 27 ns in LOS and 35.9 ns in NLOS. At 16.95 GHz, RMS DS values are 28.1 ns and 31.7 ns, respectively. The omnidirectional RMS DS, synthesized from directional PDPs \cite{Sun2015gc}, shows higher values: 63.5 ns (LOS) and 74.1 ns (NLOS) at 6.75 GHz, and 46.5 ns (LOS) and 65.8 ns (NLOS) at 16.95 GHz.
	
	\begin{table}[!t]
		\centering
		\caption{RMS Delay Spread Characteristics from 6.75 GHz to 142 GHz for LOS and NLOS Conditions.}
		\label{tab:RMS_DS}
        \resizebox{0.45\textwidth}{!}{
		\begin{tabular}{|@{\hspace{5 pt}}p{1.2 cm}@{\hspace{5 pt}}|@{\hspace{5 pt}}p{0.5 cm}@{\hspace{5 pt}}|@{\hspace{2 pt}}p{0.9 cm}@{\hspace{2 pt}}|@{\hspace{5 pt}}p{0.5 cm}@{\hspace{8 pt}}|@{\hspace{2 pt}}p{0.9 cm}@{\hspace{2 pt}}|@{\hspace{5 pt}}p{0.5 cm}@{\hspace{5 pt}}|@{\hspace{5 pt}}p{0.5 cm}@{\hspace{5 pt}}|@{\hspace{5 pt}}p{0.5 cm}@{\hspace{5 pt}}|}
			\hline
			\textbf{Frequency (GHz)} & \textbf{6.75} & \textbf{6.75 (3GPP)} & \textbf{16.95} & \textbf{16.95 (3GPP)} & \textbf{28\cite{Xing2021_Inicl}} & \textbf{73\cite{Shakya2024TAP}} & \textbf{142\cite{Shakya2024TAP}} \\ \hline
			\multicolumn{8}{|c|}{Dir RMS DS}     \\ \hline
			LOS \newline $\mathbb{E}(\cdot)$ (ns) & 29.1    &--          & 28.1   &--            & 19.3            & 14.7             & 3.6              \\ \hline
			NLOS \newline $\mathbb{E}(\cdot)$ (ns) & 35.6    &--          & 31.7   &--            & 25.7            & 23.4            & 9.2              \\ \hline
			\multicolumn{8}{|c|}{Omni RMS DS}          \\ \hline
			LOS \newline $\mathbb{E}(\cdot)$ (ns) & 62.8     &52.7          & 46.5   &42.9            & 26.7            & 16.4             & 15.4              \\ \hline
			NLOS \newline $\mathbb{E}(\cdot)$ (ns)& 75.6   &111.1           & 65.8   &96.65            & 46.3            & 50.8            & 27.3              \\ \hline
			
		\end{tabular}}
		\vspace{-1\baselineskip}
		% \vspace{-5 pt}
	\end{table}

	\renewcommand{\arraystretch}{1.05}
\begin{table*}[ht]
	\centering
		% \vspace{10pt}
 \color{black}
        \caption{Point-data table for site-specific (UMi) large scale spatio-temporal statistics \cite{Ted2025icc, Ying2025tcom} }
        % \vspace{-5 pt}
	\begin{tabular}{p{0.6 cm}p{0.5 cm}p{0.5 cm}p{0.69 cm}p{0.6 cm}p{0.69 cm}p{0.69 cm}p{0.69 cm}p{0.69 cm}p{0.69 cm}p{0.69 cm}p{0.69 cm}p{0.69 cm}p{0.6 cm}p{0.6 cm}p{0.6 cm}p{0.6 cm}}
		\hline
		\multicolumn{1}{p{0.6 cm}}{\textbf{Freq.}} & \textbf{TX} & \textbf{RX} & \textbf{Loc.} & \multicolumn{1}{p{0.6 cm}}{\textbf{TR Sep.}} & \multicolumn{1}{p{0.69 cm}}{\textbf{Omni Abs. PL (V-V)}} & \multicolumn{1}{p{0.69 cm}}{\textbf{Omni Abs. PL (V-H)}} & \multicolumn{1}{p{0.69 cm}}{\textbf{Mean Dir. DS}} & \multicolumn{1}{p{0.69 cm}}{\textbf{Omni DS}} & \multicolumn{1}{p{0.69 cm}}{\textbf{Mean Lobe ASA}} & \multicolumn{1}{p{0.69 cm}}{\textbf{Omni ASA}} & \multicolumn{1}{p{0.69 cm}}{\textbf{Mean Lobe ASD}} & \multicolumn{1}{p{0.69 cm}}{\textbf{Omni ASD}} & \multicolumn{1}{p{0.6 cm}}{\textbf{Mean Lobe ZSA}} & \multicolumn{1}{p{0.6 cm}}{\textbf{Omni ZSA}} & \multicolumn{1}{p{0.6 cm}}{\textbf{Mean Lobe ZSD}} & \multicolumn{1}{p{0.6 cm}}{\textbf{Omni ZSD}} \\
		\hline
		\text{[GHz]}& & & &[m]&[dB]&[dB]&[ns]&[ns]&[$^\circ$]&[$^\circ$]&[$^\circ$]&[$^\circ$]&[$^\circ$]&[$^\circ$]&[$^\circ$]&[$^\circ$] \\ 
		\hline
			\multirow{20}{*}{6.75} & \multirow{7}{*}{TX1} & RX1   & LOS   & 40    & 74.63 & 84.77 & 41.1  & 29.3  & 23.5  & 23.5  & 19.3  & 24.6  & 9.0   & 12.2  & 2.8   & 8.6 \\
			&       & RX2   & LOS   & 100   & 82.85 & 106.7 & 46.5  & 60.5  & 14.5  & 14.5  & 13.9  & 22.3  & 18.2  & 18.2  & 1.7   & 8.5 \\
			&       & RX3   & LOS   & 193   & 86.89 & 104.6 & 23.5  & 121.3 & 6.4   & 6.4   & 6.3   & 65.1  & 6.4   & 19.8  & 2.7   & 12.9 \\
			&       & RX4   & NLOS  & 560   & 110   & 136.6 & 2.2   & 66.4  & 10.7  & 10.7  & 12.9  & 22.0  & 4.4   & 15.4  & 2.3   & 9.5 \\
			&       & RX5   & NLOS  & 880   & 124.8 &  --     & 0.0*   & 0.0*   & 5.8   & 5.8   & 5.8   & 5.8   & 5.8   & 13.1  & 5.8   & 5.8 \\
			&       & RX6   & NLOS  & 1000  & \multicolumn{12}{c}{Outage} \\
			&       & RX7   & LOS   & 424   & 100.2  & 121.8 & 37.5  & 184.6 & 10.4  & 10.4  & 10.5  & 81.6  & 4.1   & 15.4  & 7.4   & 13.3 \\
			\cline{2-17}
			& \multirow{3}{*}{TX2} & RX1   & LOS   & 58    & 80.74 & 95.28 & 25.2  & 60.9  & 18.8  & 63.0  & 21.9  & 24.0  & 21.5  & 20.8  & 3.3   & 9.8 \\
			&       & RX2   & NLOS  & 51.5  & 88.93 & 107.3 & 34.6  & 117.7 & 13.6  & 68.7  & 18.9  & 35.5  & 17.4  & 16.5  & 3.3   & 9.8 \\
			&       & RX3   & NLOS  & 100.9 & 101.2 & 125.5 & 71.4  & 105.5 & 18.4  & 18.4  & 18.1  & 65.9  & 2.6   & 14.3  & 5.6   & 13.9 \\
			\cline{2-17}
			& \multirow{3}{*}{TX3} & RX1   & NLOS  & 71    & 101   & 117.7 & 60.4  & 171.4 & 34.0  & 34.0  & 34.4  & 36.4  & 11.7  & 11.7  & 7.8   & 7.8 \\
			&       & RX2   & NLOS  & 45    & 88.92 & 109.2 & 47.5  & 55.7  & 23.8  & 23.8  & 39.8  & 47.8  & 9.7   & 12.9  & 8.8   & 8.6 \\
			&       & RX3   & NLOS  & 125   & 100.6 & 125.3 & 24.6  & 66.5  & 18.3  & 55.7  & 33.4  & 37.1  & 9.9   & 15.9  & 10.0  & 9.8 \\
			\cline{2-17}
			& \multirow{4}{*}{TX4} & RX1   & NLOS  & 88.7  & 97.7  & 117.1 & 35.4  & 49.5  & 16.1  & 16.1  & 69.2  & 70.2  & 2.3   & 13.8  & 7.8   & 8.1 \\
			&       & RX2   & LOS   & 120   & 87.25 & 110.2 & 70.3  & 22.8  & 10.5  & 10.5  & 10.6  & 13.4  & 4.3   & 14.8  & 4.4   & 11.4 \\
			&       & RX3   & NLOS  & 216   & \multicolumn{12}{c}{Outage} \\
			&       & RX4   & NLOS  & 185   & 124.5 &  --     & 31.0  & 34.0  & 12.6  & 12.6  & 16.5  & 16.5  & 4.5   & 4.5   & 3.2   & 3.2 \\
			\cline{2-17}
			& \multirow{3}{*}{TX5} & RX1   & LOS   & 52    & 81.98 & 102.8 & 35.1  & 26.8  & 10.8  & 10.8  & 11.2  & 19.1  & 4.3   & 15.5  & 4.4   & 10.5 \\
			&       & RX2   & NLOS  & 170   & 101.5 & 125.1 & 31.5  & 46.2  & 24.2  & 24.2  & 10.5  & 32.6  & 3.8   & 12.6  & 4.2   & 12.1 \\
			&       & RX3   & NLOS  & 141   & 106.3 & 129.9 & 40.0  & 89.4  & 46.5  & 46.5  & 22.0  & 75.8  & 2.3   & 10.1  & 7.1   & 10.7 \\
			\hline
			\hline
			\multirow{20}{*}{16.95} & \multirow{7}{*}{TX1} & RX1   & LOS   & 40    & 82.14 & 113.4 & 35.9  & 34.0  & 11.8  & 11.8  & 10.9  & 15.2  & 5.2   & 7.3   & 0.9   & 4.0 \\
			&       & RX2   & LOS   & 100   & 91.39 & 121.4 & 28.7  & 56.2  & 7.2   & 7.2   & 8.2   & 18.7  & 9.6   & 9.6   & 2.4   & 5.6 \\
			&       & RX3   & LOS   & 193   & 96.04 & 124.1 & 26.5  & 118.5 & 6.6   & 67.0  & 7.3   & 57.1  & 7.6   & 12.4  & 1.8   & 5.6 \\
			&       & RX4   & NLOS  & 560   & 112.8 & 138.9 & 14.7  & 168.6 & 10.1  & 10.1  & 6.9   & 78.0  & 2.7   & 8.0   & 4.2   & 7.7 \\
			&       & RX5   & NLOS  & 880   & 130.1 &   --    & 0.0*   & 0.0*   & 5.8   & 5.8   & 5.8   & 5.8   & 5.4   & 5.4   & 5.4   & 5.4 \\
			&       & RX6   & NLOS  & 1000  & \multicolumn{12}{c}{Outage} \\
			&       & RX7   & LOS   & 410$^\dagger$   & 113.1 & 138.4 & 65.9  & 97.9  & 5.8   & 5.8   & 10.4  & 47.8  & 5.4   & 11.5  & 7.3   & 9.0 \\
			\cline{2-17}
			& \multirow{3}{*}{TX2} & RX1   & LOS   & 58    & 86.3  & 106.8 & 46.2  & 16.9  & 16.6  & 16.6  & 6.7   & 22.0  & 3.2   & 8.3   & 4.1   & 8.9 \\
			&       & RX2   & NLOS  & 51.5  & 89.61 & 118.8 & 94.5  & 37.3  & 8.0   & 8.0   & 7.5   & 25.1  & 7.0   & 9.6   & 4.0   & 7.3 \\
			&       & RX3   & NLOS  & 100.9 & 115.2 & 138.1 & 104.9 & 146.2 & 19.5  & 19.5  & 9.3   & 20.4  & 6.3   & 7.6   & 6.1   & 9.7 \\
			\cline{2-17}
			& \multirow{3}{*}{TX3} & RX1   & NLOS  & 71    & 114.4 & 141.1 & 83.7  & 180.8 & 20.9  & 28.0  & 10.4  & 28.4  & 2.9   & 8.1   & 4.7   & 8.4 \\
			&       & RX2   & NLOS  & 45    & 103.9 & 128.6 & 9.7   & 34.3  & 5.6   & 19.8  & 10.5  & 31.4  & 8.8   & 12.1  & 7.1   & 9.8 \\
			&       & RX3   & NLOS  & 125   & 111.4  & 123.8 & 25.6  & 64.5  & 14.4  & 33.5  & 15.0  & 23.5  & 11.4  & 13.6  & 11.9  & 11.5 \\
			\cline{2-17}
			& \multirow{4}{*}{TX4} & RX1   & NLOS  & 88.7  & 102.9 & 126.2 & 19.2  & 31.9  & 7.7   & 33.9   & 7.3   & 80.4  & 4.1   & 9.9   & 6.3   & 8.1 \\
			&       & RX2   & LOS   & 120   & 95.46 & 123.8 & 34.7  & 22.1  & 4.8   & 46.8  & 6.3   & 10.9  & 8.0   & 9.8   & 3.6   & 7.6 \\
			&       & RX3   & NLOS  & 216   & \multicolumn{12}{c}{Outage} \\
			&       & RX4   & NLOS  & 185   & 129.5 &  --     & 18.0  & 18.0  & 11.9  & 11.9  & 9.9   & 24.2  & 4.8   & 4.8   & 5.6   & 5.2 \\
			\cline{2-17}
			& \multirow{3}{*}{TX5} & RX1   & LOS   & 52    & 91.11 & 115.9 & 86.4  & 23.6  & 5.8   & 5.8   & 5.8   & 12.0  & 5.4   & 11.6  & 5.4   & 8.9 \\
			&       & RX2   & NLOS  & 170   & 111.7 &  --     & 17.5  & 28.1  & 25.6  & 25.6  & 9.1   & 18.5  & 9.3   & 9.3   & 4.2   & 7.2 \\
			&       & RX3   & NLOS  & 141   & 122.5 &  --     & 28.9  & 60.5  & 28.3  & 28.3  & 9.5   & 47.7  & 5.5   & 7.3   & 3.5   & 4.2 \\
			\hline
			\multicolumn{17}{l}{\footnotesize *only have one MPC from a single direction at this location.\,\, $^\dagger$RX at 16.95 GHz moved 14 m closer due to inaccessibility of exact location at 6.75 GHz.}
		\end{tabular}%
		\label{tab:LSPs}%
		\vspace{-5 pt}
	\end{table*}%
	\medskip
	\renewcommand{\arraystretch}{1}

	A clear decreasing trend in RMS DS is observed as frequency increases, evident in both directional and omnidirectional measurements for LOS and NLOS scenarios. This trend suggests reduced multipath dispersion at higher frequencies, consistent with findings at mmWave frequencies \cite{Xing2021_Inicl,Shakya2024TAP}.
	The mean values are evaluated in log-scale and the expectations are calculated assuming underlying log-normal distribution for direct comparison with 3GPP. Our measurements at 6.75 GHz and 16.95 GHz show slightly higher RMS DS values compared to the 3GPP models in LOS scenarios (62.8 ns NYU measured vs. 52.7 ns 3GPP predicted at 6.75 GHz), while measured NLOS RMS DS is lower compared to the 3GPP model results (75.6 ns NYU measured vs. 111.1 ns 3GPP predicted at 6.75 GHz). \textcolor{black}{The 3GPP model assumes generalized UMi environments with statistically averaged scattering density~\cite{3GPPTR38901}. However, the measured outdoor urban environment in MetroTech Commons open-square in Brooklyn featured fewer dynamic scatterers in NLOS paths (e.g., sparse foliage, and open plazas) compared to the dense urban scenarios modeled by 3GPP. This reduced scattering leads to fewer resolvable MPCs, resulting in lower NLOS RMS DS measurements. Conversely, in LOS conditions, the reflective far-spaced building surfaces surrounding open square environment (Fig. \ref{fig:ODMap}) introduces additional late arriving MPCs with several tens of nanoseconds delay, resulting in a measured RMS DS greater than 3GPP models (Table \ref{tab:RMS_DS}).}

\subsection{Angular Spread Measurement Results}\vspace{-4 pt}
 Table \ref{tab:AS} presents the omnidirectional RMS AS statistics, including azimuth spread of arrival (ASA) and azimuth spread of departure (ASD), evaluated from the measurements. 
 \textcolor{black}{The AS is commonly evaluated using the 3GPP method \cite{3GPPTR38901} (circular standard deviation of MPCs) or Fleury's method (second central moment of the PAS) \cite{Ying2025tcom}. A comprehensive comparison between the two approaches in computing AS is presented in \cite{Ying2025tcom}.} Following the 3GPP conventions, the AS is characterized using log-normal distributions parameterized by $\mu$ and $\sigma$, enabling direct comparison with 3GPP model predictions. To ensure statistical comparability with typical UMi scenarios, measurements outside the open square, along the urban street canyon in Fig. \ref{fig:ODMap} (b), beyond 180 m were excluded for analysis as these distant locations exhibited significantly reduced AS.

\setlength{\tabcolsep}{10pt} 
\begin{table*}[t!]
	\centering
	%\footnotesize
	\caption{Angular Spread Characteristics measured by NYU WIRELESS at 6.75 GHz and 16.95 GHz in LOS and NLOS (T-R Sep.$<$180 m) for the UMi environment and Comparison to 3GPP Models \cite{3GPPTR38901,Shakya2025wcnc,Ying2025tcom}.}
	% \vspace{-5 pt}
	\label{tab:AS}
	\renewcommand{\arraystretch}{1.2} % More space between rows
	\begin{tabular}{|@{\hspace{5 pt}}P@{\hspace{20 pt}}|e|c|p{1.5 cm}@{\hspace{2 pt}}|x|x|c|p{0.6 cm}|x|x|}
		\specialrule{1.2 pt}{0 pt}{0 pt}
		
		\multirow{2}{*}{\textbf{Metric}} & \multirow{2}{*}{\textbf{Condition}} & \multicolumn{4}{c|}{\textbf{6.75 GHz}} & \multicolumn{4}{c|}{\textbf{16.95 GHz}} \\
		\cline{3-10}
		& & \textbf{NYU\cite{Shakya2025wcnc}} & \textbf{3GPP\cite{3GPPTR38901}} & $\mathbb{E}(\textbf{NYU})$ & $\mathbb{E}(\textbf{3GPP})$ & \textbf{NYU} & \textbf{3GPP} & $\mathbb{E}(\textbf{NYU})$ & $\mathbb{E}(\textbf{3GPP})$ \\
		\specialrule{1 pt}{0 pt}{0 pt}
		
		\multirow{4}{*}{\shortstack{Omni RMS $\lg_\text{ASA}$\\  $=\log_{10}$($\text{ASA}/1^{\circ}$)}}
		& $\mu_\text{lgASA}^\text{LOS}$& 1.28 & \centering 1.66 & \multirow{2}{*}{21.44$^\circ$} & \multirow{2}{*}{50.36$^\circ$} & 1.12 & 1.63 & \multirow{2}{*}{15.30$^\circ$} & \multirow{2}{*}{47.31$^\circ$} \\ \cline{2-3} \cline{4-4} \cline{7-8} 
		& $\sigma_\text{lgASA}^\text{LOS}$ & 0.32 & \centering 0.29 &  &  & 0.36 & 0.30 &  &  \\ \cline{2-10}
		& $\mu_\text{lgASA}^\text{NLOS}$ & 1.50 & \centering 1.74 & \multirow{2}{*}{33.61$^\circ$} & \multirow{2}{*}{62.78$^\circ$} & 1.36 & 1.71 & \multirow{2}{*}{23.99$^\circ$} & \multirow{2}{*}{59.54$^\circ$} \\ \cline{2-3} \cline{4-4} \cline{7-8} 
		& $\sigma_\text{lgASA}^\text{NLOS}$ & 0.23 & \centering 0.34 &  &  & 0.20 & 0.36 &  &  \\ \specialrule{1 pt}{0 pt}{0 pt}

		\multirow{4}{*}{\shortstack{Omni RMS $\lg_\text{ASD}$\\=$\log_{10}$($\text{ASD}/1^{\circ}$)}}
		& $\mu_\text{lgASD}^\text{LOS}$ & 1.31 & \centering 1.16 & \multirow{2}{*}{20.70$^\circ$} & \multirow{2}{*}{17.54$^\circ$} & 1.18 & 1.15 & \multirow{2}{*}{15.43$^\circ$} & \multirow{2}{*}{17.14$^\circ$} \\ \cline{2-3} \cline{4-4} \cline{7-8} 
		& $\sigma_\text{lgASD}^\text{LOS}$ & 0.11 & \centering 0.41 &  &  & 0.13 & 0.41 &  &  \\ \cline{2-10}
		& $\mu_\text{lgASD}^{\text{NLOS}}$ & 1.67 & \centering 1.32 & \multirow{2}{*}{48.00$^\circ$} & \multirow{2}{*}{25.85$^\circ$} & 1.49 & 1.24 & \multirow{2}{*}{32.51$^\circ$} & \multirow{2}{*}{22.41$^\circ$} \\ \cline{2-3} \cline{4-4} \cline{7-8} 
		& $\sigma_\text{lgASD}^\text{NLOS}$ & 0.15 & \centering 0.43 &  &  & 0.21 & 0.47 &  &  \\ 
		\specialrule{1.2 pt}{0 pt}{0 pt}
		
	\end{tabular}
	\vspace{-10 pt}
\end{table*}
\renewcommand{\arraystretch}{1}

\begin{figure}
	\centering%
	\subfloat[]{%
		\centering
		\includegraphics[width=43mm]{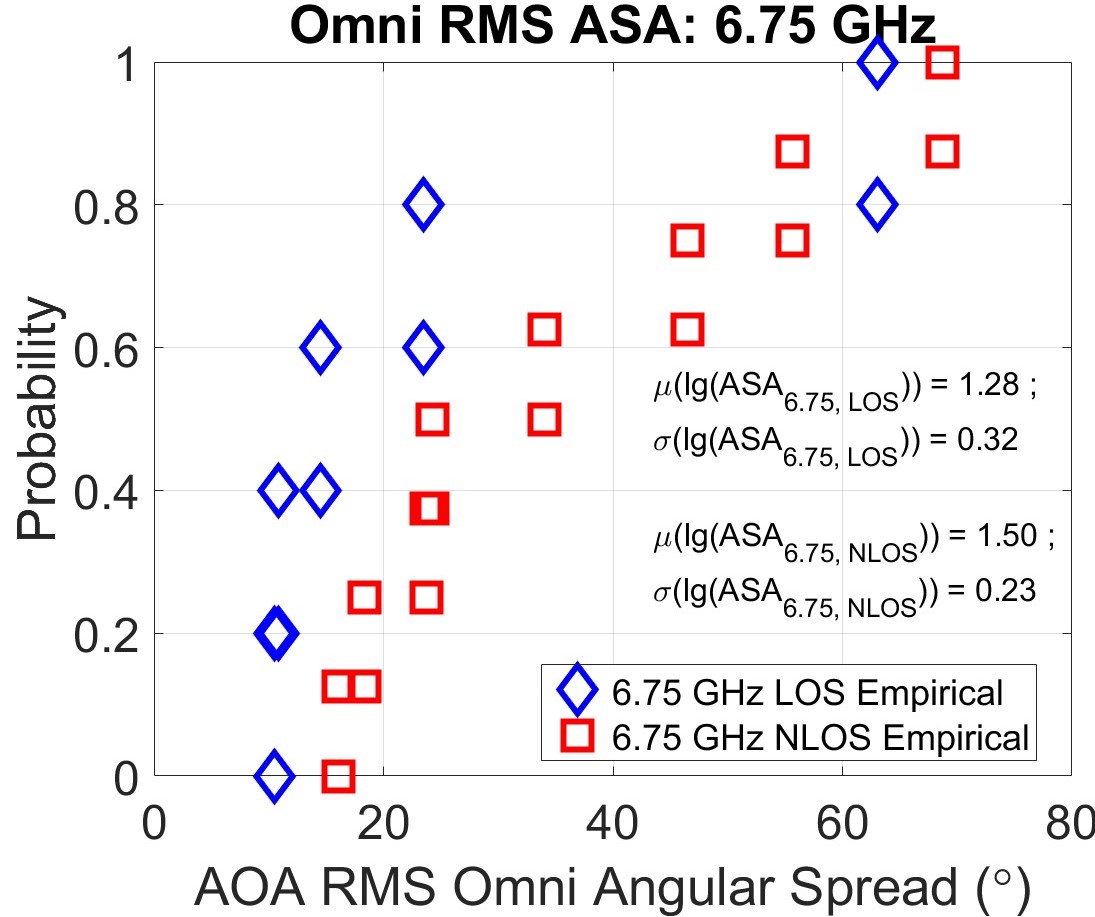}
	}
	\hspace{-6 pt}
	\subfloat[]{%
		\centering
		\includegraphics[width=43mm]{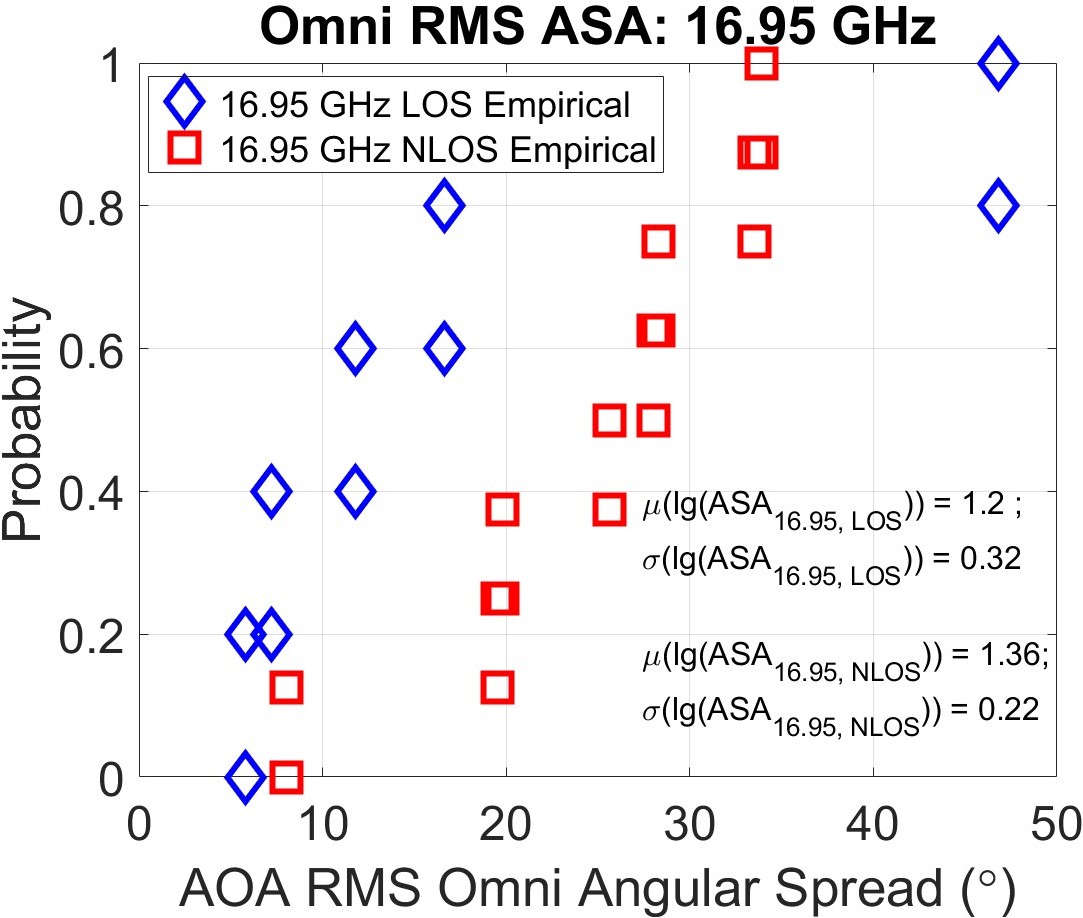}
	}%
	%[2.6mm]
    \vspace{-5 pt}
	\caption{UMi omnidirectional ASA CDF: (a) 6.75 GHz; (b) 16.95 GHz. [T-R separation: 40--180 m]}
	\label{fig:ASA_D}
	\vspace{-20 pt}
\end{figure}

The CDFs of ASA for both frequencies are presented in Fig. \ref{fig:ASA_D}. At 6.75 GHz, the measured omnidirectional RMS ASA shows expected values of 21.44$^\circ$ in LOS and 33.61$^\circ$ in NLOS conditions, which are substantially lower than the 3GPP model predictions of 50.36$^\circ$ and 62.78$^\circ$, respectively. The 16.95 GHz measurements also yield much smaller spreads, with expected values of 15.30$^\circ$ in LOS and 23.99$^\circ$ in NLOS, compared to 3GPP model values of 47.31$^\circ$ and 59.54$^\circ$. For the omni RMS ASD, our measurements at 6.75 GHz reveal larger spreads than the 3GPP model, with expected values of 20.70$^\circ$ in LOS and 48.00$^\circ$ in NLOS, compared to 3GPP values of 17.54$^\circ$ and 25.85$^\circ$, respectively. Similar trends are observed at 16.95 GHz, where measured expected values are 15.43$^\circ$ in LOS and 32.51$^\circ$ in NLOS, versus 3GPP predictions of 17.14$^\circ$ and 22.41$^\circ$. The measurements consistently show frequency-dependent behavior in AS, with higher frequencies exhibiting smaller AS. The observations suggest that current 3GPP models need revision to closely model the observed AS characteristics. Location-specific spatio-temporal statistics given in point-data format are summarized in Table \ref{tab:LSPs} and follows the detailed approach presented for indoor hotspot in \cite{Ted2025icc}.

\section{Conclusion}
\vspace{-2 pt}
Extensive outdoor propagation measurements at 6.75 GHz and 16.95 GHz conducted in Downtown Brooklyn, USA were presented encompassing seven LOS and 13 NLOS locations with two outages. The measurements reveal lower PLEs in these bands compared to mmWave frequencies, suggesting better coverage potential for urban deployments. Additionally, a decreasing trend in RMS DS was observed with increasing frequency. RMS AS also showed a decreasing trend at higher frequencies. Compared to 3GPP models, the RMS ASA and NLOS RMS DS were found to be consistently lower at both 6.75 GHz and 16.95 GHz. The results can provide valuable insights for facilitating air-interface and network design for 5G and 6G communications, e.g., for beam management, where beam alignment is a key factor especially for higher number of elements at gNB and different AS values reported.

	\section*{Acknowledgment}
    \vspace{-2 pt}
	Authors thank Prof. Sundeep Rangan and Prof. Marco Mezzavilla for support with licensing.
    \vspace{-2 pt}
	
	\bibliographystyle{IEEEtran}
	\bibliography{references}
	
\end{document}